\documentclass[aps,prl,showpacs,twocolumn,floatfix]{revtex4-1}

\usepackage{amsmath,amssymb,epsfig,amsbsy,xspace,empheq}
\usepackage{color}
\usepackage{subfigure}
\usepackage{multirow}
\usepackage{hyperref}

\hypersetup{colorlinks=true,
linkcolor=blue,
citecolor=blue,
urlcolor=blue
}

\newcommand{\be}{\begin{equation}}
\newcommand{\ee}{\end{equation}}

\newcommand{\ZZq}{\mathbb{Z}_q}

\begin{document}

\title{Critical exponents can be different on the two sides of a transition: A generic mechanism}

\author{F. L\'eonard, B. Delamotte }

\affiliation{Laboratoire de Physique Th\'eorique de la Mati\`ere Condens\'ee, 
CNRS UMR 7600, UPMC-Sorbonne Universit\'es, 75252 Paris Cedex 05, France
}


\begin{abstract}
We present models where $\gamma_+$ and $\gamma_-$, the exponents of the  susceptibility
in the high and low temperature phases, are generically different. In these models,
continuous symmetries  are explicitly broken down by discrete anisotropies that 
are irrelevant in the renormalization-group sense. The $\ZZq$-invariant models are 
the simplest examples for two-component order parameters ($N=2$) and the icosahedral symmetry
for $N=3$. We compute accurately  $\gamma_+ -\gamma_-$
as well as the ratio $\nu/\nu'$ of the exponents of the two correlation lengths present for $T<T_c$.

\end{abstract}

\date{\today}

\maketitle

The question of the equality of the critical exponents on the two sides of a second order phase transition has, apparently, not been raised for decades.
The general renormalization group (RG) argument ``showing'' their equality goes as follows: Correlation functions are regular in the presence
of an external field, which is sufficient to go continuously from one phase to the other.
Moreover, if these functions satisfy the same RG equations above and below the critical temperature $T_c$, the same should hold true for the scaling behavior
of quantities such as the susceptibility, the correlation length or the specific heat. 
Since the renormalization properties of a theory are identical in its symmetric and spontaneously broken
phases,  it follows that the critical exponents are identical in both phases 
(see for instance \cite{zinn2002quantum,kardar2007statistical,le2012phenomenes}). This is indeed what  happens
generically.

To the best of our knowledge, Nelson \cite{PhysRevB.13.2222} was the first to 
propose a counterexample based on the $O(2)$ model
in dimension $d=3$, to which is added either a 
cubic (CA) \cite{PhysRevB.13.2222,PhysRevA.10.435,Amit1982207,PhysRevB.8.4270,0301-0015-6-11-006,0022-3719-6-8-007,0022-3719-7-12-003,PhysRevB.11.478,PhysRevB.61.15136} 
or hexagonal (HA) anisotropy \cite{PhysRevB.13.2222,PhysRevA.10.435,Amit1982207}. These anisotropies
are taken into account in the Ginzburg-Landau hamiltonian  by  terms of order 4 and 6, respectively, which 
 are  irrelevant in the RG sense at the transition. The corresponding fixed point is thus $O(2)$ symmetric. However, Nelson
argued \cite{PhysRevB.13.2222} that they are dangerously irrelevant \cite{FisherDangerously,Amit1982207}
in the low-temperature phase  and that
they, therefore, induce a modification of the exponent $\gamma_-$ of the susceptibility. 
A  rather counterintuitive result is that the
difference $\gamma_+ - \gamma_-$ is larger for HA than for CA, whereas HA is ``more irrelevant" than CA. A detailed
study of the literature shows that, up until now, 
this striking result has been completely ignored.

Because of its relationship with either deconfined quantum critical points \cite{PhysRevLett.99.207203} or pyrochlore \cite{PhysRevB.89.140403} 
and the possible existence of two distinct phase
transitions \cite{miyashita97}, the three-dimensional XY model with HA (and more generally the $\ZZq$-invariant models) has been studied again
\cite{PhysRevB.61.3430,PhysRevB.91.174417}. Although only one transition
has been found, the $\ZZq$ models were shown to exhibit 
two correlation lengths below $T_c$, $\xi$ and $\xi'$,  that scale with two different critical exponents, $\nu$ and $\nu'$. 
All authors agree that $\nu/\nu'$ depends on the scaling dimension of the irrelevant HA term, but there
are no less than three different scaling relations predicting this ratio, as well as several values obtained by Monte Carlo 
simulations \cite{PhysRevB.61.3430,PhysRevLett.99.207203,PhysRevB.91.174417}.

In this Letter, we present a mechanism, valid not only for the XY case, 
to generate different critical exponents above and below $T_c$.
 The mechanism relies on the possibility of explicitly breaking a continuous symmetry down to a discrete
one by terms that are irrelevant in the RG sense. In addition to the $\ZZq$-invariant models, we build an example for Heisenberg spins
showing icosahedral symmetry. Using the nonperturbative renormalization group (NPRG), it is---contrary to perturbation
theory---easy to show that the exponents $\gamma_+$ and $\gamma_-$ are generically different for these models and easy to compute them,
as well as $\nu/\nu'$, accurately. 
Our approach allows us to completely clarify the physics of these models. 

Let us discuss the general idea underlying the difference between  $\gamma_+$ and $\gamma_-$. 
For concreteness,  we consider a XY or Heisenberg model described by an O$(N)$-invariant Hamiltonian ($H_{{\rm O}(N)}$),
to which is added a discrete anisotropy term $\tau(x)$,
$H=H_{{\rm O}(N)}+ \lambda_{\rm an}\int_x  \tau(x)$.
We assume that $\tau(x)$ is irrelevant in $d=3$.  The fixed point (FP) describing 
the phase transition is therefore O$(N)$ symmetric 
($\lambda_{\rm an}^{\rm FP}=0$). 
If this term were irrelevant in the ordinary sense---that is,
could be neglected  ($\lambda_{\rm an}=0$)---the model would be identical to the 
O$(N)$ model. It is important to remember that, in this case,
not only the transverse  ($\chi_{\rm T}$) 
but also the longitudinal  ($\chi_{\rm L}$) susceptibilities
 diverge for all $T<T_c$ because of the  Goldstone
modes \cite{patashinskii1973longitudinal,PhysRevB.7.1967,PhysRevE.83.031120}. However, since the symmetry 
is discrete when $\lambda_{\rm an}\neq0$, there are no Goldstone modes 
and the susceptibilities cannot diverge  for $T<T_c$.
Thus, $\chi_{\rm T}^{-1}$ vanishes only at $T_c$, and its scaling with $\Delta T=T-T_c$ 
obviously depends on the way  $\lambda_{\rm an}$ scales to zero close to the fixed point.
Since this scaling is given by the scaling dimension of $\tau(x)$, the exponent $\gamma_{\rm T}$ defined by 
$\chi_{\rm T}^{-1}\sim (T_c-T)^{\gamma_{\rm T}}$ for $T<T_c$ cannot be equal to  $\gamma_+$.
This is why $\tau(x)$ is said to be dangerously irrelevant for $T<T_c$ . 
The same holds true for $\chi_{\rm L}^{-1}$. 

Let us now give two examples, with $N=2$ and 3, of the kind of anisotropy that produces a difference between $\gamma_+$ and $\gamma_{T,L}$. 
We choose discrete subgroups $G$ of either $O(2)$ or $O(3)$. These subgroups must satisfy two constraints. First,
in order to have only one phase transition, there must exist only one invariant quadratic polynomial of $G$.
Therefore, it must be $\varphi_i\varphi_i$, $i=1,\cdots, N$ as in the O$(N)$ model.
Second, the interaction term that explicitly breaks
the O$(N)$ symmetry must be irrelevant. A term of order 4 can be irrelevant compared to the O$(N)$-invariant term
$(\varphi_i\varphi_i)^2$. For $N=2$ and  $d=3$, this is the case, for instance,  of the term $\varphi_1^4+\varphi_2^4$ of cubic anisotropy. For reasons
that are explained below, this kind of terms, being ``weakly irrelevant", induces only small differences
between $\gamma_+$ and $\gamma_{T,L}$ that, moreover, require very large systems to be observable. We therefore prefer
to consider terms that are ``strongly irrelevant",  because they are of degrees higher  than 4.

For $N=2$, all the $\ZZq$-invariant models ($q$-state clock models)
with $q>4$ satisfy the two  conditions above  because the first invariant polynomial in $\varphi_i$ that is not $O(2)$-symmetric
is of degree $q$. For instance, for  $\mathbb{Z}_6$, this invariant reads : 
$\tau=6\varphi_1^5\varphi_2+6\varphi_1\varphi_2^5 - 20\varphi_1^3\varphi_2^3$ \cite{Amit1982207} \footnote{We notice an error in Eq.~(4.1) 
of \cite{PhysRevB.13.2222} about the hamiltonian ${\cal H}_w$
corresponding to HA.}.

For $N=3$, the situation is more constrained because  only the icosahedral group satisfies the two conditions
above \cite{refId0}. For all the other discrete subgroups of $O(3)$, the first invariant polynomials that are not $O(3)$ symmetric are of degree
4 and, therefore, are at best only weakly irrelevant. For the icosahedral symmetry, the first non-$O(3)$-symmetric 
invariant polynomial  is of degree 6  and reads \cite{cummins1988polynomial,refId0}: 
\begin{equation}
 \begin{aligned}
 \tau=& (4 \Phi - 2) (\varphi_1^2 - \varphi_2^2) ( \varphi_2^2 - \varphi_3^2) ( \varphi_3^2 - \varphi_1^2) + 22 (\varphi_1 \varphi_2 \varphi_3)^2\\
 &+ (\varphi_1^4 + \varphi_2^4 + \varphi_3^4) (\varphi_1^2 + \varphi_2^2 + \varphi_3^2),
 \end{aligned}
\end{equation}
where $\Phi$ is the golden ratio \footnote{Two other independent invariant polynomials, of degree ten and fifteen, exist for this group.}.

The NPRG is based on Wilson's idea of integrating fluctuations step by step 
 \cite{PhysRevB.4.3174}. In its modern
version, it is implemented on the Gibbs free energy $\Gamma$ \cite{Wetterich199390,Ellwanger1993,MorrisTR1994}.
A one-parameter family of models indexed by a scale $k$ is thus defined such that only the rapid
fluctuations, with wavenumbers $\vert q\vert>k$, are summed over in the partition function ${\cal Z}_k$. 
The decoupling of the slow modes ($\vert q\vert<k$) in ${\cal Z}_k$ is performed by adding to 
the original hamiltonian $H$ a quadratic (``mass-like'') term which is nonvanishing only for these modes:
\begin{equation}
 {\cal Z}_k[\pmb{J}]= \int D\pmb{\varphi} \exp(-H[\pmb{\varphi}]-\Delta H_k[\pmb{\varphi}]+ \pmb{J\cdot\varphi})
\end{equation}
with  $\Delta H_k[\pmb{\varphi}]= \frac{1}{2} \int_q R_k(q^2) \varphi_i(q)\varphi_i(-q)$---where, for instance, 
$R_k(q^2)=Z_k(k^2-q^2)\theta(k^2-q^2)$, with $\theta$
the step function and $Z_k$ the field renormalization constant---and $\pmb{J\cdot\varphi}=\int_x J_i(x) \varphi_i(x)$. 
The coarse-grained Gibbs free energy $\Gamma_k[\pmb{\phi}]$
is defined as  the (slightly modified) Legendre transform of $\log  {\cal Z}_k[J_i]$:
\begin{equation}
\label{legendre}
 \Gamma_k[\pmb{\phi}]+\log  {\cal Z}_k[\pmb{J}]= \pmb{J\cdot\phi}-\frac 1 2 \int_q R_k(q^2) \phi_i(q)\phi_i(-q)
 \end{equation}
 where $\phi_i(x)$ is the thermal  average of $\varphi_i(x)$.
When $k$ is  of the order of the inverse lattice
spacing $\Lambda$, all fluctuations in ${\cal Z}_k$ are frozen by the $R_k$ term 
and the mean-field approximation becomes exact. With the definition (\ref{legendre}), this implies
that $\Gamma_{k=\Lambda}[\pmb{\phi}]= H[\pmb{\phi}]$ \cite{Wetterich199390}. Since $R_{k=0}(q^2)\equiv 0$,  $\Gamma_{k=0}[\pmb{\phi}]=\Gamma[\pmb{\phi}]$ 
and is, thus, the free energy that we want to compute. 

The exact flow equation of $\Gamma_k$ reads \cite{Wetterich199390} (see Supplemental Material for more details):
\begin{equation}
\label{flow}
\partial_t\Gamma_k[\pmb{\phi}]=\frac 1 2 {\rm Tr} [\partial_t R_k(q^2) (\Gamma_k^{(2)}[q,-q;\pmb{\phi}]+R_k(q))^{-1}]
\end{equation}
where $t=\log(k/\Lambda)$, ${\rm Tr}$ stands for an integral over $q$ and a trace over group indices, and $\Gamma_k^{(2)}[q,-q;\pmb{\phi}]$ is
the matrix of the Fourier transforms of the second functional derivatives of $\Gamma_k[\pmb{\phi}]$ with respect to $\phi_i(x)$
and $\phi_j(y)$. 

Since it is impossible to solve
Eq.~(\ref{flow}) exactly, we must make use of approximations. 
To capture the critical physics, the simplest nonperturbative  approximation is the derivative expansion \cite{Tetradis1994541,Berges2002223}.
We use the (improved) lowest order, the local potential approximation prime (LPA'), which consists of retaining only a potential term in
$\Gamma_k[\pmb{\phi}]$ together with a field renormalization constant $Z_k$ in front of the kinetic term \cite{Tetradis1994541,Berges2002223},
\begin{equation}
  \Gamma_k^{\rm LPA'}[\pmb{\phi}]= \int_{x}\left\{ \frac{1}{2} Z_k[{\nabla} \phi_i(x)]^2 + 
 U_k(\phi_i(x))\right\}.
 \label{LPA}
\end{equation}

The running potential $U_k$ is defined by $\Omega\, U_k(\phi_i)=\Gamma_k[\phi_i]$,
 where the fields $\phi_i$ are constant and $\Omega$ is the volume
of the system. Its flow is obtained from Eq.~(\ref{flow}) where $\Gamma_k^{(2)}$ is computed from 
(\ref{LPA}) and is then  evaluated in a constant field configuration $\phi_i$.

On top of the LPA',  $U_k(\phi_i)$ can be expanded around one of its running minima $\phi_{i,k}^{\rm min}$,
 which corresponds, at $k=0$, to the stable state of the system when $J_i=0$,
\begin{equation}
\label{field-expansion}
 U_k = \frac{u_{20,k}}{2} (\rho-\kappa_k)^2+ u_{01,k} \tau+\frac {u_{30,k}}{3!}(\rho-\kappa_k)^3+ \cdots
\end{equation}
with $\rho=\phi_i\phi_i/2$ and $\kappa_k=\phi_{i,k}^{\rm min}\phi_{i,k}^{\rm min}/2$.  
The flows of  $u_{mn,k}$ and $\kappa_k$ are obtained from that of $U_k$ by acting  with $\partial_t$ on both sides of their definition,
\begin{equation}
 u_{mn,k}=\frac{\partial^{m+n}U_k}{\partial\rho^m\partial\tau^n}{\vert_{\rho=\kappa_k,\tau=0}}\, ,\ \ \ 
 \frac{\partial U_k}{\partial\rho}{\vert_{\rho=\kappa_k,\tau=0}}=0.
\end{equation}

\begin{figure}[top]
 \centering
 \subfigure[]{\includegraphics[scale=0.30]{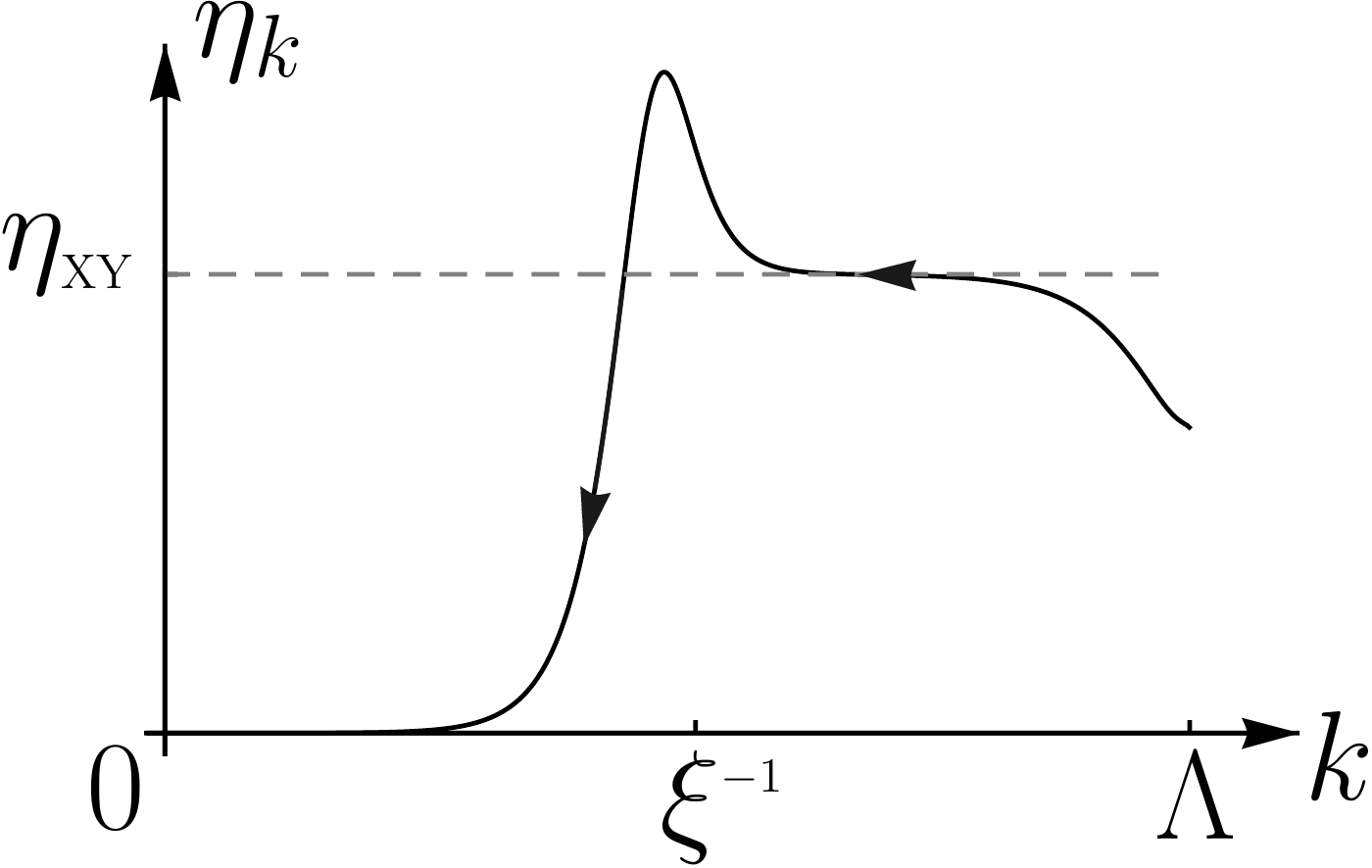}\label{fig1a}}
 \subfigure[]{\includegraphics[scale=0.30]{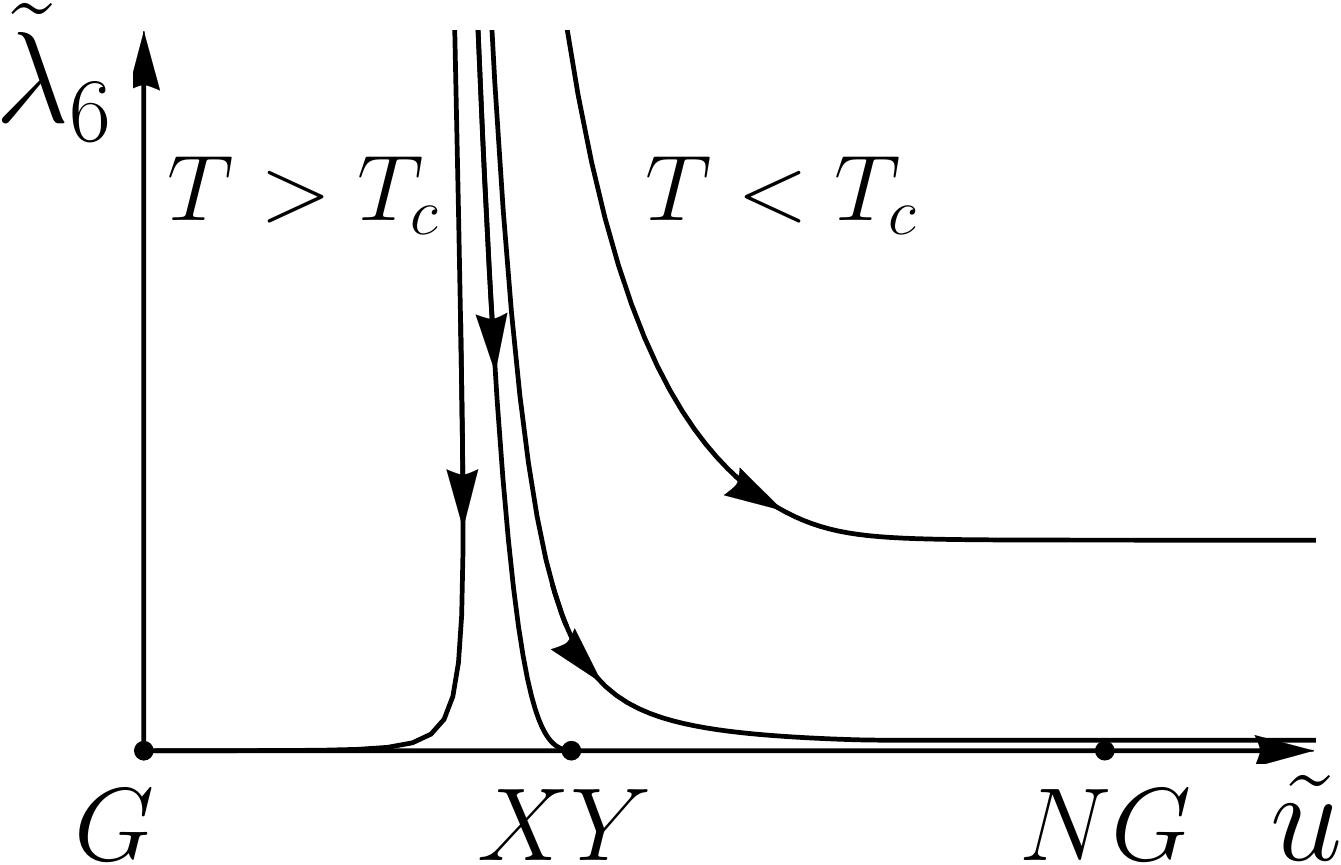}\label{fig1b}}
 \subfigure[]{\includegraphics[scale=0.30]{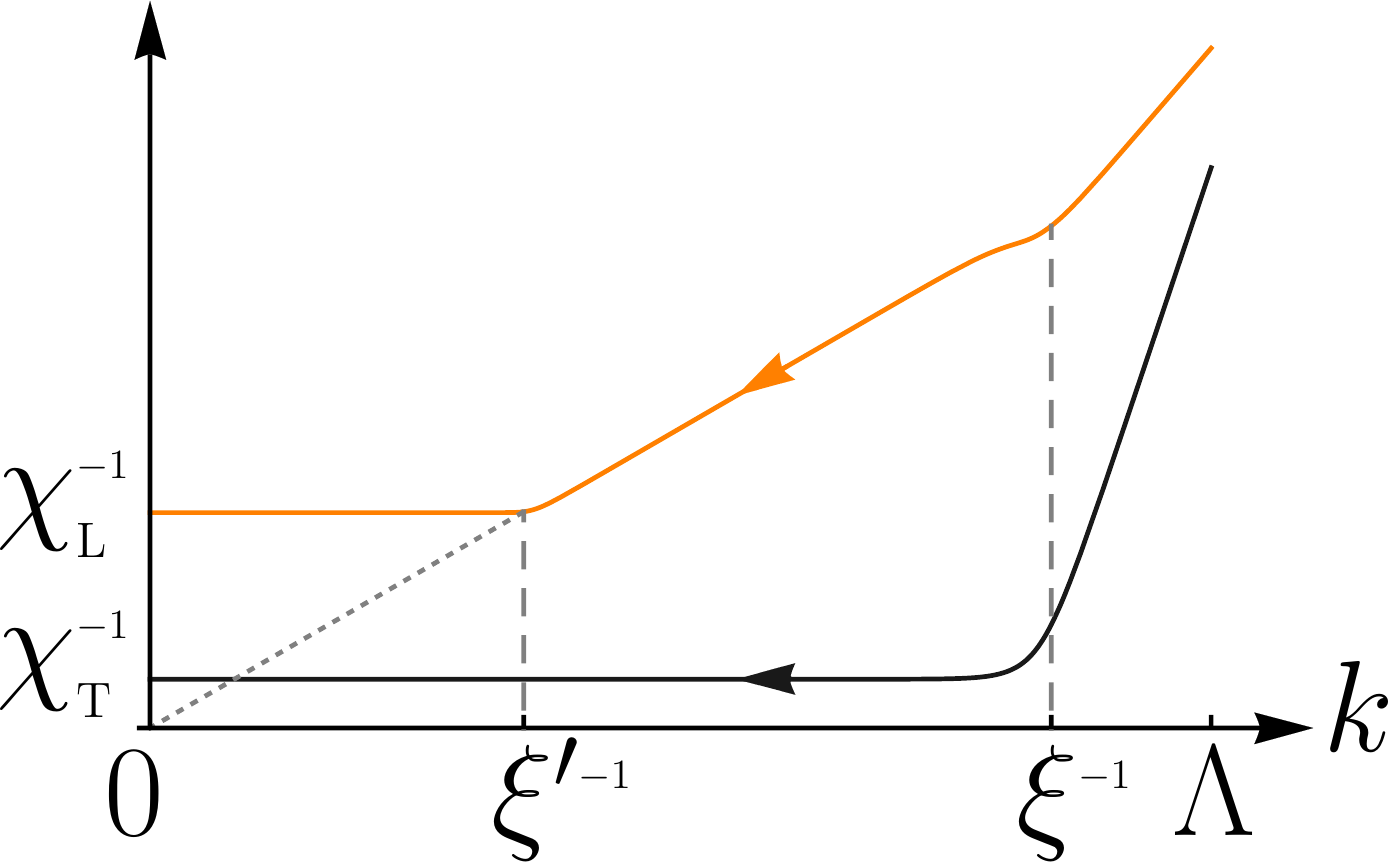}\label{fig1c}}
 \subfigure[]{\includegraphics[scale=0.30]{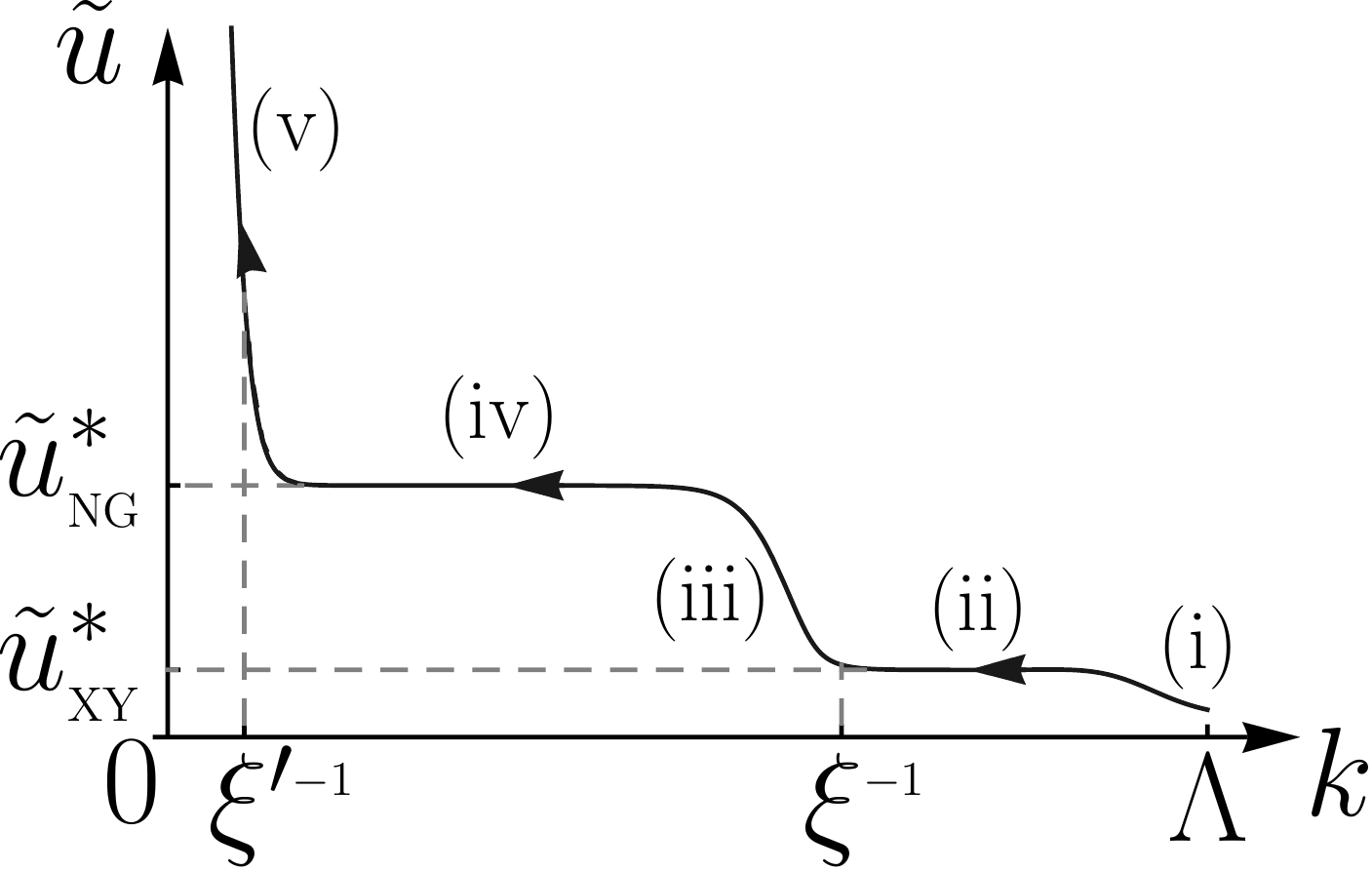}\label{fig1d}}
\caption{XY model with HA in $d=3$.
(a) Flow for $T<T_c$ of the running anomalous dimension. (b) Projections of RG trajectories in the $\tilde u - \tilde\lambda_6$ plane.  G, XY and NG are the Gaussian, critical O(2) 
and low-temperature (Nambu-Goldstone) fixed points. (c) Flows for $T<T_c$ of the inverse transverse and longitudinal susceptibilities. 
The dashed line represents the flow of $\chi_L^{-1}$ in the pure O(2) model. (d) Flow of $\tilde u$ for $T<T_c$ showing its five different regimes. 
The two plateaus correspond to the vicinity of the XY and NG FP.}
\label{figflow}
\end{figure}

We have performed the calculations up to order 12 in the field-expansion equation~(\ref{field-expansion}). 
However, for the sake of simplicity, we present below 
the RG flow obtained for the XY model with HA within the simplest ansatz that includes only $u_{20,k}$ and $u_{01,k}$, which
we call $u$ and $\lambda_6$ (we omit the $k$ index in the following to alleviate the notation). In this case,
$\phi_{i,k}^{\rm min}$ corresponds to one of the six minima of $U_k$; we call the direction pointing 
towards $\phi_{i,k}^{\rm min}$ longitudinal and the perpendicular direction transverse. Once diagonalized, 
$\Gamma_{ij,k}^{(2)}$ splits as usual into  the two inverse longitudinal and transverse propagators that each depend 
on a (running) ``mass" $2 u \kappa$ and $18 \lambda_6 \kappa^2$, which we call $m_{\rm L}$ and $m_{\rm T}$. 
Finding RG fixed points requires us to work with dimensionless and renormalized variables.
We thus define $\tilde\kappa =  K_d^{-1} Z_k k^{2-d}\kappa$,
$\tilde u=K_d Z_k^{-2} k^{d-4}u$, and $\tilde\lambda_6= K_d^{2} Z_k^{-3} k^{2d-6}\lambda_6$, where
$K_d^{-1} = d \,\Gamma({d}/{2})2^{d-1}\pi^\frac{d}{2}$ has been included for convenience. The 
flow equations read (see Supplemental Material for more details):
\begin{subequations}
 \label{flowLPA}
 \begin{empheq}{align}
 \displaystyle{  \partial_t \tilde\kappa} = & \displaystyle{(2-d-\eta_k)\tilde\kappa + \left( \frac{1}{2} + \frac{18 \tilde\kappa \tilde\lambda_6}{\tilde u} \right) I_2(\tilde m_{T}^2)}\nonumber\\
					   &+ \frac{3}{2} I_2(\tilde m_{L}^2)\\
 \displaystyle{   \partial_t \tilde u} = & \displaystyle{ (d-4+2\eta_k)\tilde u - 18  \tilde\lambda_6 I_2(\tilde m_{T}^2)+ 9 \tilde{u}^2 I_3(\tilde m_{L}^2)} \nonumber\\
                                         & \displaystyle{ + ( \tilde u + 36 \tilde\kappa \tilde\lambda_6 )^2 I_3(\tilde m_{T}^2)}\\
 \displaystyle{  \partial_t \tilde\lambda_6 }  = & \displaystyle{(2d-6+3\eta_k)\tilde\lambda_6} \nonumber\\
                                               &\displaystyle{ +15  \tilde\lambda_6 ( \tilde u + 6 \tilde\kappa \tilde\lambda_6 )\frac{ I_2(\tilde m_{T}^2) - I_2(\tilde m_{L}^2)}{\tilde m_{L}^2-\tilde m_{T}^2}}     
\end{empheq}
\end{subequations}
with  $I_n(x)= 2 (1 + x)^{-n}(1-{\eta_k}/{(d+2)})$,
 $\tilde{m}_L^2=2 \tilde u\tilde\kappa$, $\tilde{m}_T^2=18 \tilde\lambda_6 \tilde\kappa^2$.
The running anomalous dimension is defined by $\eta_k=- \partial_t \log Z_k$ which tends, at criticality, to the anomalous dimension $\eta$ for $k\to~0$ \cite{Tetradis1994541,Berges2002223}. 
We show its flow in Fig.~\ref{fig1a}.

The flow equations (\ref{flowLPA}) are very simple, though they
are nonperturbative.
They show two crucial features. First, they take automatically into account the role of the masses $m_{\rm L,T}$ and their decoupling: 
as long as $k\gtrsim m_{\rm L,T}$, that is, $1\gtrsim\tilde m_{\rm L,T}$, the contributions coming from 
$ I_n(\tilde m_{\rm L}^2)$ 
(resp.  $ I_n(\tilde m_{\rm T}^2)$) are non vanishing; they become negligible when $k\ll m_{\rm L,T}$,
and the longitudinal (resp. transverse)  mode is then said to decouple from the flow.
Second, once generalized to arbitrary $N$, they reproduce the low-$T$ expansion of 
the O($N$) nonlinear-sigma model at one loop in $d=2+\epsilon$ (when $\lambda_6=0$) and are also one-loop exact in $d=4-\epsilon$ \cite{Berges2002223,PhysRevB.69.134413}.
Gathering all these  properties in a single set of flow equations is out of reach of the usual perturbative 
expansions.

Equations (\ref{flowLPA}) admit three fixed points (with noninfinite  couplings 
$\tilde u$ and $\tilde\lambda_6$) as shown in Fig.~\ref{figflow}. The NG FP has coordinates 
$\tilde u_{\rm NG}^*=2-d/2, \tilde\lambda_{6,{\rm NG}}^*=0, \tilde\kappa_{\rm NG}^*=\infty$.
By integrating Eqs.~(\ref{flowLPA}) in $d=3$ with different initial conditions, we find the RG trajectories 
shown in Fig.~\ref{fig1b}. For $T<T_c$, $\lambda_6$ and $\kappa$ reach fixed, nonvanishing values for  $k\lesssim\xi^{-1}$.
This is why the dimenionful inverse transverse susceptibility $\chi^{-1}_T$ stops running beyond this scale, see Fig.~\ref{fig1c}.
The dimensionless analogues of these couplings, $\tilde\lambda_6$ and $\tilde\kappa$,
keep running according to their canonical dimension. Thus, for $k\ll\xi^{-1}$,
\begin{equation}
\begin{array}{l}
 \displaystyle{\tilde\kappa(k) \sim  \tilde\kappa(\xi^{-1})\, (k\xi)^{2-d}
\,  \sim\,  \tilde\kappa^*_{\rm XY}\,  (k\xi)^{2-d}}\\
 \\
 \displaystyle{\tilde\lambda_6(k)\sim\tilde\lambda_6(\xi^{-1})\, (k\xi)^{2d-6}
 \sim\tilde\lambda_6^{\rm in}\, \xi^{-\vert y_6\vert}\, (k\xi)^{2d-6}}
 \end{array}
 \label{flow-beyond-xi}
\end{equation}
where $y_6$ is the scaling exponent of $\tilde\lambda_6$ and $\tilde\lambda_6^{\rm in}$ is the initial value of $\tilde\lambda_6$, that is, its value at the scale of the inverse lattice spacing.
As for $\tilde u$, we find in $d=3$, for $T$ slightly below $T_c$, five different regimes  represented on Fig.~\ref{fig1d}.
In region (v), $\tilde u$ diverges and $u$ reaches a finite value.

The values of $k$ at which the RG flow departs, respectively, from the XY and NG FP  define two length scales
called $\xi$ and $\xi'$, see Fig.~\ref{figflow}. The first one, $\xi$, is the Josephson
length (the correlation length
of the amplitude mode) of the pure O(2) model \cite{Josephson1966608} [the anisotropy plays no role in part (ii) of the flow, see Fig.~\ref{fig1d}, when
$\xi$ is large  because $\tilde\lambda_6(k\sim\xi^{-1})\ll1$]. 
As long as the RG trajectory remains close to  NG, the flows of $\tilde u$ and $\tilde\kappa$
 remain similar to the flows of the O(2) model in the low-$T$ phase [region (iv) in Fig.~\ref{fig1d}], and $\chi_L^{-1}$ 
 decreases as it does in
 the pure O(2) model where the fluctuations of the Goldstone modes make it vanish, see Fig.~\ref{fig1c}. 
 However, because $\tilde\lambda_6\neq0$, the flow departs from the NG FP \cite{PhysRevB.61.3430,PhysRevLett.99.207203}; $u(k)$ stops flowing,
 as does $\chi_L^{-1}$, see Fig.~\ref{fig1c}. 
 This occurs when $\tilde{m}^2_T(k={\xi'}^{-1})\simeq1$, which defines ${\xi'}$. This correlation length diverges at $T_c$ according to
 $\xi'\sim \Delta T^{-\nu'}$, by definition of $\nu'$.
 Using the definition of $\tilde{m}^2_T$ and Eqs.~(\ref{flow-beyond-xi}), it is now straightforward to show that
 \begin{equation}
 \label{nuprime}
  \nu'=\nu(1+\vert y_6\vert/2).
 \end{equation}
This relation  was already obtained in \cite{PhysRevB.91.174417}, although we do not fully agree with its derivation 
\footnote{Contrary to what is shown in  Fig.~1 of \cite{PhysRevB.91.174417}, we find only 
three fixed points with $\tilde\lambda_6<\infty$.
Moreover, in $d=3$,   $\tilde\lambda_6$ 
cannot be of order 1 for $T\to T_c$ and $k\to0$.}. The derivation above shows that $\nu'$ is universal, contrary
to what is suggested in \cite{PhysRevB.89.140403}.

 The scaling relations among  exponents 
 are readily derived from the discussion above and the usual scaling behavior of the potential,
 \begin{equation}
 \label{scaling-U}
  U= s^d U\left( s^{-1/\nu}\Delta T, \tilde{u}(s),\tilde\lambda_6(s),\cdots, \tilde\phi Z(s)^{1/2}\right)
 \end{equation}
where $s$ is a rescaling factor, the dots stand for the infinite set of irrelevant couplings, and $Z(s)$ is 
the field renormalization factor. $Z(s)\sim s^{-d+2-\eta_{\rm XY}}$ for $s\sim (\xi \Lambda)^{-1}$, 
where $\eta_{\rm XY}$ is the anomalous dimension at the XY FP, 
and $Z(s)\sim s^{-d+2}$ for $s\ll(\xi \Lambda)^{-1}$ since $\eta_k$ vanishes away from it (in particular 
at the NG FP), see Fig.~\ref{fig1a}.
In the low-$T$ phase,
by taking two derivatives of (\ref{scaling-U}) with respect to $\tilde\phi$ in either in the transverse or longitudinal directions,
and then taking $\tilde\phi$ at the minimum of $U$, we obtain
\begin{eqnarray}
%\begin{array}{l}
 &&\chi_T^{-1}\propto s^d Z(s) \tilde\lambda_6 (s) \tilde\kappa^2(s)\label{susceptibilities1}\\
 &&\chi_L^{-1}\propto s^d Z(s) \tilde u(s) \tilde\kappa(s).\label{susceptibilities2}
% \end{array}
 \label{susceptibilities}
\end{eqnarray}
By taking $s\sim (\xi \Lambda)^{-1}$ in Eq.~(\ref{susceptibilities1}) we obtain: $\gamma_T=\gamma_+ + \nu \vert y_6\vert$, and
by taking $s\sim (\xi'\Lambda)^{-1}$ and using Eqs.~(\ref{flow-beyond-xi}), (\ref{nuprime}) we obtain 
$\gamma_L=\gamma_+ + (4-d)\nu \vert y_6\vert/2$ \footnote{Notice that we  disagree with Eq.~(4.3)
of \cite{PhysRevB.13.2222} although we agree with Eq.~(4.12) of the same paper.}.
Notice that the scaling relations derived  above for $\mathbb{Z}_6$ are generically valid.  

We have computed $y_q$ up to order 12 in field-expansion equation (\ref{field-expansion}) to obtain converged results, see Table~\ref{table}.
We observe that our value of $\nu'$  for $\mathbb{Z}_6$ is very close to the one deduced from the scaling law Eq.~(\ref{nuprime}) and Monte Carlo
simulations in \cite{PhysRevB.91.174417}. This validates our approach.
We find,  of course, that $\vert y_q\vert$ increases with $q$; we thus deduce 
the rather counterintuitive result that the more irrelevant  the anisotropy
term, the larger the difference between  $\gamma_+$ and $\gamma_{T,L}$. However, since $\xi'$ diverges
extremely rapidly close to $T_c^-$ for large $\vert y_q\vert$, it must be difficult to observe the scaling behavior
of $\chi_L$ in a finite-size system for ``large" values of $q$. As for $\chi_T$, its measurement should not be 
 more difficult  than in the pure O(2) model. 
Reciprocally, if $\vert y_q\vert$ is too small, the transient regime before reaching 
the XY FP  is  very  large [region (i) in Fig.~\ref{fig1d}], and, thus, the corrections to scaling are also large; this spoils
an accurate determination of the leading scaling behavior in finite-size systems. This is probably the case of CA in $d=3$, where 
we find $\vert y_4\vert=0.042$ [at six loops $\vert y_4\vert=0.103(8)$ \cite{PhysRevB.61.15136}]. 

\begin{table}[t]
 \centering
 \begin{ruledtabular}
 \begin{tabular}{lllll}
   \multicolumn{1}{c}{Symmetry} & \multicolumn{1}{l}{$\mathbb{Z}_4$} & \multicolumn{1}{l}{$\mathbb{Z}_5$} & \multicolumn{1}{l}{$\mathbb{Z}_6$} &
\multicolumn{1}{l}{$\mathbb{Z}_{10}$}  \\
  \hline
  \multirow{2}{1.5cm}{\centering $\nu'$} & $0.71$ & $1.06$& $1.44$ & $3.84$  \\
  &$0.72${ \cite{PhysRevLett.99.207203}} & $1.05${ \cite{PhysRevLett.99.207203}}&  $1.6${ \cite{PhysRevLett.99.207203}} & $2.8${ \cite{PhysRevLett.99.207203}}\\
  & & & $1.45${ \cite{PhysRevB.91.174417}} & \\
  \hline
  \multirow{2}{1.5cm}{\centering $\gamma_T-\gamma_+$} & $0.029$ & $0.74$ & $1.49$ & $6.29$ \\
   & $0.06${ \cite{PhysRevB.61.15136}} & & $1.58${ \cite{PhysRevB.13.2222}} & \\
 \end{tabular}
 \end{ruledtabular}
 \caption{ Critical exponents  in $d=3$ for the XY, $\mathbb{Z}_q$  models. 
 In both cases, the first row corresponds to our results. For the icosahedral symmetry ($N=3$), we find at order
 eight in field expansion Eq.~(\ref{field-expansion}):
 $\nu'=1.51$ and $\gamma_T-\gamma_+=1.54$ .}
 \label{table}
\end{table}

We have presented a general mechanism leading to a large and measurable difference between critical exponents in
the high- and low-$T$ phases and a theoretical approach to compute them. For the XY case, we have resolved the existing 
discrepancies between
the results obtained in $\ZZq$-invariant models \cite{PhysRevB.61.3430,PhysRevLett.99.207203,PhysRevB.89.140403,PhysRevB.91.174417}.
Let us also emphasize that layered decagonal quasicrystals \cite{KraposhinV.S.2011,Rau20041330} showing
tenfold anisotropies and  XY spin systems with HA  \cite{PhysRevB.13.2222} exist,
which would enable a direct measurement of $\gamma_{T,L}-\gamma_+$ and $\nu'$.
Another very interesting challenge is the possibility of measuring susceptibilities in Heisenberg systems
with icosahedral anisotropy, possibly in quasicrystals. We recall that, for $N=3$, there are problaby many other anisotropies that are
dangerously irrelevant; this likely would lead to differences between $\gamma_{T,L}$ and $\gamma_+$ that are  smaller than in the icosadrehal
case, but that are possibly also measurable.
Finally, it would be extremely interesting to investigate the two-dimensional 
\cite{PhysRevB.16.1217,PhysRevLett.40.561,0305-4470-13-4-037,PhysRevB.26.6201,Landau19831115,PhysRevB.33.437,0305-4470-24-1-033,PhysRevB.68.104409,PhysRevLett.112.155702,PhysRevB.90.205109}
case with the NPRG approach. 
At the price of avoiding any field truncation and  working in at least the second order of the derivative
expansion \cite{Ballhausen2004144,Morris1995139,PhysRevE.90.062105}, this is reachable. We leave it for future work.

\begin{acknowledgments}
 We acknowledge D.R. Nelson for pointing out to us the existence of quasicrystals with tenfold anisotropies, 
 P. Sindzingre for discussions and clarifications about numerical issues, A. Ran\c{c}on for pointing out to us some important references, and
 F. Benitez, H. Chat\'e, N. Dupuis, F. Rose, M. Tissier and N. Wschebor for discussions 
 and suggestions regarding the manuscript.
\end{acknowledgments}

% \bibliography{jabref.bib}
% \bibliographystyle{h-physrev.bst}

\end{document}

\title{Supplemental material : Critical exponents can be different on the two sides of a transition: A generic mechanism}

\author{F. L\'eonard, B. Delamotte }

\affiliation{Laboratoire de Physique Th\'eorique de la Mati\`ere Condens\'ee, 
CNRS UMR 7600, UPMC-Sorbonne Universit\'es, 75252 Paris Cedex 05, France
}

\maketitle

% \section{Detailed calculation of the flow equation of the Gibbs free energy}
\subsection{Definitions and conventions}
We define :
\begin{equation}
 \displaystyle{\int_x = \int d^d x} \hspace{0.5cm}\text{  and  } \hspace{0.5cm}\displaystyle{\int_q = \int \frac{d^d q}{(2\pi)^d}},
\end{equation}
where $x$ and $q$ are respectively position and wavenumber vectors and
$ \pmb{J\cdot\varphi}=\int_x J_i(x) \varphi_i(x)$.
The coarse-grained partition function is defined by :
\begin{equation}
 {\cal Z}_k[\pmb{J}]= \int D\pmb{\varphi} \exp(-H[\pmb{\varphi}]-\Delta H_k[\pmb{\varphi}]+ \pmb{J\cdot\varphi})
 \label{eq:partitionfunction}
\end{equation}
where $\Delta H_k[\pmb{\varphi}]=\frac{1}{2}\int_{x,y} R_k(x-y) \varphi_i(x)\varphi_i(y)$, and the Helmoltz $W_k$ and modified Gibbs $\Gamma_k$ free energies by :
\begin{subequations}
 \begin{empheq}{align}
   W_k[\pmb{J}] = &\log  {\cal Z}_k[\pmb{J}]\label{eq:GammaWa}\\
   \Gamma_k[\pmb{\phi}]+W_k[\pmb{J}]=& \:\pmb{J\cdot\phi}-\Delta H_k[\pmb{\phi}]\label{eq:GammaWb}.
 \end{empheq}
 \label{eq:GammaW}
\end{subequations}
From \eqref{eq:GammaW} one finds :
\begin{subequations}
 \begin{empheq}{align}
  &\displaystyle{\frac{\delta W_k[\pmb{J}]}{\delta J_i(x)} = \langle\varphi_i(x)\rangle = \phi_i(x)}\\
  &\displaystyle{\frac{\delta \Gamma_k[\pmb{\phi}]}{\delta \phi_i(x)} = J_i(x) - \int_y R_k(x-y) \phi_i(y)}.
 \end{empheq}
 \label{eq:derivlegendre}
\end{subequations}
We also define :
\begin{subequations}
\begin{empheq}{align}
  W_k^{(2)}[x,y;\pmb{J}]_{i,j} &= \displaystyle{\frac{\delta^2 W_k}{\delta J_i(x)\delta J_j(y)}}\\
  \Gamma_k^{(2)}[x,y;\pmb{\phi}]_{i,j} &= \displaystyle{\frac{\delta^2 \Gamma_k}{\delta \phi_i(x)\delta \phi_j(y)}}
\end{empheq}
\end{subequations}
and the Fourier transform :
\begin{equation}
 \Gamma_k^{(2)}[q,q';\pmb{\phi}]_{i,j} = \int_{x,y} e^{i\left( x q + x'q'\right)}\Gamma_k^{(2)}[x,x';\pmb{\phi}]_{i,j}\:.
\end{equation}

\subsection{Flow equations for the coarse-grained free energies}
From \eqref{eq:partitionfunction} and \eqref{eq:GammaWa}, we obtain :
\begin{eqnarray}
 \label{eq:flowWk}
  \partial_k W_{k|\pmb{J}} = &-& \frac{1}{2}\int_{x,y} \partial_k R_k(x-y) \left( W_k^{(2)}(x,y)_{i,i}\right.\nonumber\\
   &+& \left. \frac{\delta W_k}{\delta J_i(x)}\frac{\delta W_k}{\delta J_i(y)} \right).
\end{eqnarray}
 Using : 
 \begin{equation}
  \displaystyle{\partial_{k|\pmb{J}} = \partial_{k|\pmb{\phi}} + \int_x \partial_k \phi_i(x)_{k|\pmb{J}}\: \frac{\delta }{\delta \phi_i(x)}},
 \end{equation}
 and Eqs.~\eqref{eq:GammaWb} and \eqref{eq:flowWk}, one finds :
\begin{equation}
 \label{eq:flowGW}
 \partial_{k\vert\pmb{\phi}}\,\Gamma_k[\pmb{\phi}]=\frac 1 2 {\rm Tr} [\:\partial_k R_k(x-y) W_k^{(2)}[x,y;\pmb{J}]\:]
\end{equation}
where ${\rm Tr}$ stands for an integral over $x$ and $y$ and a trace over group indices.
 From Eqs.~\eqref{eq:derivlegendre}, one finds :
 \begin{equation}
 \label{eq:WinvG}
  W_k^{(2)}[x,y;\pmb{J}] = \left( \Gamma_k^{(2)}[x,y;\pmb{\phi}] + R_k(x-y) \right)^{-1},
 \end{equation}
 where the inverse has to be understood in the operator sense.
 Substituting \eqref{eq:WinvG} in \eqref{eq:flowGW} we obtain the flow of $\Gamma_k[\pmb{\phi}]$ in Fourier space (see Eq.~(4) in the article):
 \begin{equation}
 \label{eq:flowGamma}
 \partial_t\Gamma_k[\pmb{\phi}]=\frac 1 2 {\rm Tr} [\:\partial_t R_k(q^2) (\Gamma_k^{(2)}[q,-q;\pmb{\phi}]+R_k(q))^{-1}\:],
 \end{equation}
where $t=\log(k/\Lambda)$.

\subsection{Derivation of the flow equations (8)}
The LPA' consists in truncating $\Gamma_k$ as :
\begin{eqnarray}
  \Gamma_k^{\rm LPA'}[\pmb{\phi}]&=& \int_{x} \left[ \frac{1}{2} Z_k ({\nabla} \pmb{\phi}(x))^2 +  U_k(\pmb{\phi}(x))\right].
 \label{eq:LPA}
\end{eqnarray}
The running potential $U_k$ is defined by: $\Omega\, U_k(\phi_i)=\Gamma_k[\phi_i]$ 
where the fields $\phi_i$ are constant and $\Omega$ is the volume
of the system. Its flow is obtained from Eq.~\eqref{eq:flowGamma} where $\Gamma_k^{(2)}$ is computed from 
\eqref{eq:LPA} and then  evaluated in a constant field configuration $\phi_i$.
Until $U_k(\pmb{\phi})$ is specified the following calculations apply to all $\mathbb{Z}_q$ anisotropies.
In the following, $\Gamma_k$ means $\Gamma_k^{\rm LPA'}$ in order to alleviate the notation.

With the {\it ansatz}~\eqref{eq:LPA} we obtain for the hexagonal anisotropic XY-model,
\begin{widetext}
\begin{equation}
\begin{split}
\Gamma_k^{(2)}[q;\pmb{\phi}] + R_k(q) = \left(
 \begin{array}{cc}
  R_k(q) + Z_k\:q^2 + U_k^{(2,0)}(\pmb{\phi}) & U_k^{(1,1)}(\pmb{\phi})\\
  U_k^{(1,1)}(\pmb{\phi}) & R_k(q) + Z_k\:q^2 + U_k^{(0,2)}(\pmb{\phi})
 \end{array}
 \right) \: \text{ where } \:U_k^{(i,j)}(\pmb{\phi}) = \displaystyle{\frac{\partial^{i+j} U_k(\pmb{\phi})}{\partial \phi_1^i \phi_2^j}}.
 \label{eq:matrixGamma2}
\end{split}
\end{equation}
\end{widetext} 
To compute the flow of $U_k$  we have to invert the matrix in Eq.~\eqref{eq:matrixGamma2} which requires to diagonalize it.
Its eigenvalues are :
 \begin{subequations}
  \begin{empheq}{align}
   &R_k(q) + Z_k\:q^2 + m_{k,L}^2(\phi_1,\phi_2)\\
   &R_k(q) + Z_k\:q^2 + m_{k,T}^2(\phi_1,\phi_2)
  \end{empheq}
 \end{subequations}
 where $m_{k,L,T}^2$ are given below.
 Once evaluated in a constant field configuration $\phi_i$, Eq.~\eqref{eq:flowGamma} becomes :
 \begin{eqnarray}
 \label{eq:flowU}
 \partial_t U_k[\pmb{\phi}]&=&\frac 1 2 \int_q \partial_t R_k(q^2) \left(\displaystyle{\frac{1}{R_k(q) + Z_k\:q^2 + m_{k,L}^2(\phi_1,\phi_2)}}\right.\nonumber\\
 &&\left. +\displaystyle{\frac{1}{R_k(q) + Z_k\:q^2 + m_{k,T}^2(\phi_1,\phi_2)}} \right).
 \end{eqnarray}
 In order to simplify the integral in Eq.~\eqref{eq:flowU}, we use the simplest regulator :
 \begin{equation}
  \label{eq:reg}
   R_k(q^2)=Z_k(k^2-q^2)\theta(k^2-q^2),
 \end{equation}
 with $\theta$ the step function and $Z_k$ the field renormalization constant.
 Then, Eq.~\eqref{eq:flowU} reads :
 \begin{equation}
 \label{eq:flowUtheta}
  \partial_t U_k = \frac 1 2 \left[ I_1(m_{k,L}^2(\phi_1,\phi_2)) + I_1(m_{k,T}^2(\phi_1,\phi_2)) \right],
 \end{equation}
 with
 \begin{equation}
  I_n(x) = \displaystyle{ 2 (1 + x)^{-n}(1-\frac{\eta_k}{d+2}) },
 \end{equation}
 where by definition, $\eta_k=- \partial_t \log Z_k$.
 \subsection{The flow equations of the coupling constants}
 It is convenient to re-write the flow of the potential $U_k(\pmb{\phi})$ in terms of the two $\mathbb{Z}_q$-invariants.
 For the XY-model in the presence of a $\mathbb{Z}_6$-anisotropy, they read:
 \begin{subequations}
 \begin{empheq}{align}
  \rho &= \displaystyle{\frac{\phi_1^2 + \phi_2^2}{2}}\\
  \tau' &= \frac 1 6 (6\phi_1^5\phi_2+6\phi_1\phi_2^5 - 20\phi_1^3\phi_2^3)
 \end{empheq}
 \end{subequations}
 where $\rho$ is the usual $O(2)$ invariant.

 Since we expand the potential around one of its minima, it is convenient to use the freedom to change $\tau~=~\tau' + \alpha \rho^3$, with $\alpha$ real, to have a new invariant $\tau$ that vanishes at the minimum. We find :
 \begin{equation}
  \label{tau}
  \tau = \frac 1 8 (\phi_1^3 + 3\phi_1^2\varphi_2 - 3\phi_1\phi_2^2-\phi_2^3)^3.
 \end{equation}
 The masses $m_{k,L,T}^2$ become :
 \begin{widetext}
\begin{equation}
\begin{split}
 m_{k,L,T}^2(\rho,\tau) &= U_k^{(1,0)}(\rho,\tau) + \rho U_k^{(2,0)}(\rho,\tau) + 6 \tau U_k^{(1,1)}(\rho,\tau) + 9 \rho^2 U_k^{(0,1)}(\rho,\tau) + 18 \rho^2 \tau U_k^{(0,2)}(\rho,\tau)\\
 &\pm [\:9 \rho((9 \rho^3 + 20 \tau) U_k^{(0,1)}(\rho,\tau)^2 + 4 \tau U_k^{(0,1)}(\rho,\tau)(9 \rho^3 U_k^{(0,2)}(\rho,\tau) + 6 \tau U_k^{(0,2)}(\rho,\tau)+7\rho U_k^{(1,1)}(\rho,\tau))\\
 &+ \rho \tau(9 \rho^2 \tau U_k^{(0,2)}(\rho,\tau)^2 +6 \tau U_k^{(0,2)}(\rho,\tau)U_k^{(1,1)}(\rho,\tau)+2\rho U_k^{(1,1)}(\rho,\tau)^2)-6((3\rho^3-5\tau)U_k^{(0,1)}(\rho,\tau)\\
 &-2\tau(-3\rho^3 U_k^{(0,2)}(\rho,\tau)+3\tau U_k^{(0,2)}(\rho,\tau)+\rho U_k^{(1,1)}(\rho,\tau)))U_k^{(2,0)}(\rho,\tau)+\rho^2 U_k^{(2,0)}(\rho,\tau)^2\:]^{1/2},
\end{split}
\label{eq:masses}
\end{equation}
\end{widetext}
where the $\pm$  corresponds respectively to  $m_{L}^2$ and $m_{T}^2$.
 We now perform a field expansion of the potential around its running minimum $\kappa_k$:
 \begin{equation}
\label{eq:field-expansion}
 U_k = \frac{u_{20,k}}{2} (\rho-\kappa_k)^2+ u_{01,k} \tau+\frac {u_{30,k}}{3!}(\rho-\kappa_k)^3+ \cdots
\end{equation}
where $\kappa_k=\phi_{i,k}^{\rm min}\phi_{i,k}^{\rm min}/2$ with $\phi_{i,k}^{\rm min}$ one of the six minima of $U_k(\pmb{\phi})$.
The coupling constants are defined by :
\begin{equation}
\label{eq:defcoupling}
 u_{mn,k}=\frac{\partial^{m+n}U_k}{\partial\rho^m\partial\tau^n}{\vert_{\rho=\kappa_k,\tau=0}}\, ,\ \ \ 
 \frac{\partial U_k}{\partial\rho}{\vert_{\rho=\kappa_k,\tau=0}}=0.
\end{equation}
Their flows are obtained from that of $U_k$, Eq.~\eqref{eq:flowUtheta}, by acting  with $\partial_t$ on the two sides of \eqref{eq:defcoupling}.

At the minimum the masses read :
\begin{subequations}
 \begin{empheq}{align}
  &m_{k,L}^2 =  2 \kappa_k U_k^{(2,0)}(\kappa_k,0) = 2 \kappa_k u_k\\
  &m_{k,T}^2 =  18 \kappa_k^2 U_k^{(0,1)}(\kappa_k,0) = 18 \kappa_k^2 \lambda_{k,6}
 \end{empheq}
\end{subequations}
where $u_k =u_{20,k} $ and $\lambda_{k,6}=u_{01,k}$. We omit the $k$-index in the following to alleviate the notation.

Finding RG fixed points requires to work with dimensionless and renormalized variables.
We thus define: 
\begin{subequations}
 \begin{empheq}{align}
  &\tilde\rho =  K_d^{-1} Z_k k^{2-d}\rho\\
  &\tilde\tau = Z_k^3 k^{6-3d} K_d^{-3} \tau\\
  &\tilde U_k(\tilde\rho,\tilde\tau) = k^{-d} K_d^{-1} U_k(\rho,\tau)
 \end{empheq}
\end{subequations}
where $K_d^{-1} = d \,\Gamma({d}/{2})2^{d-1}\pi^\frac{d}{2}$ has been included for convenience.
Equations (8) of the article are obtained by keeping only, in the field expansion \eqref{eq:field-expansion}, the coupling constants $u$ and $\lambda_6$ and the running minimum $\kappa$.
Their dimensionless analogues read :
\begin{subequations}
 \label{eq:adim}
 \begin{empheq}{align}
  &\tilde\kappa =  K_d^{-1} Z_k k^{2-d}\kappa\\
  &\tilde u=K_d Z_k^{-2} k^{d-4}u\\
  &\tilde\lambda_6= K_d^{2} Z_k^{-3} k^{2d-6}\lambda_6.
%   &\tilde u_{30} = K_d^2 Z_k^{-3} k^{2 d - 8} u_{30}.
 \end{empheq}
\end{subequations}\\\\
The flow equations (8) of the article follow from using
\begin{equation}
 \partial_t |_{\rho,\tau} = \partial_t |_{\tilde\rho,\tilde\tau} + \partial_t \tilde\rho|_\rho \:\partial_{\tilde\rho} + \partial_t \tilde\tau|_\tau \partial_{\tilde\tau},
\end{equation}
and Eqs.~\eqref{eq:flowUtheta},\eqref{eq:masses},\eqref{eq:defcoupling} and \eqref{eq:adim}.
% \begin{widetext}
% \begin{subequations}
%  \label{eq:flowcoupling}
%  \begin{empheq}{align}
%  \displaystyle{  \partial_t \tilde\kappa} = & \displaystyle{(2-d-\eta_k)\tilde\kappa + \left( \frac{1}{2} + \frac{18 \tilde\kappa \tilde\lambda_6}{\tilde u} \right) I_2(\tilde m_{T}^2)} + \left(\frac{3}{2} + \frac{\tilde\kappa \tilde u_{30}}{\tilde u}\right) I_2(\tilde m_{L}^2)\\
%  \displaystyle{   \partial_t \tilde u} = & \displaystyle{ (d-4+2\eta_k)\tilde u - 18  \tilde\lambda_6 \left( 1 - \frac{\tilde\kappa \tilde u_{30}}{\tilde u}\right) I_2(\tilde m_{T}^2) + \tilde u_{30} \left(\frac{\tilde\kappa \tilde u_{30}}{\tilde u}-1 \right)I_2(\tilde m_{L}^2)+ \left( 3 \tilde{u} + 2 \tilde\kappa \tilde u_{30} \right)^2 I_3(\tilde m_{L}^2)} \nonumber \\
%                                          & \displaystyle{ + ( \tilde u + 36 \tilde\kappa \tilde\lambda_6 )^2 I_3(\tilde m_{T}^2)}\\
%  \displaystyle{  \partial_t \tilde\lambda_6 }  = & \displaystyle{(2d-6+3\eta_k)\tilde\lambda_6} \displaystyle{ +15  \tilde\lambda_6 ( \tilde u + 6 \tilde\kappa \tilde\lambda_6 )\frac{ I_2(\tilde m_{T}^2) - I_2(\tilde m_{L}^2)}{\tilde m_{L}^2-\tilde m_{T}^2}}     
% \end{empheq}
% \end{subequations}
% \end{widetext}
% Since $u_{30}$ does not take part in the discussion that follows in the article, we take it to zero.
\subsection{Computation of $\eta_k$}
In order to compute $\eta_k=- \partial_t \log Z_k$, we define $Z_k$ by~:
\begin{equation}
 Z_k = \lim_{q^2\to0} \displaystyle{\frac{d}{dq^2} \Gamma_k^{(2)}[q,-q;\pmb{\phi}_{min}]_{2,2}} .
\end{equation}
Its flow requires to compute the flow of $\Gamma^{(2)}_k$ :
\begin{widetext}
 \begin{equation}
  \begin{split}
   \partial_t \Gamma_k^{(2)}[q,-q;\pmb{\phi}_{min}]_{i,j} = \displaystyle{\frac{\delta \partial_t \Gamma_k[\pmb{\phi}]}{\delta \phi_i(q) \delta \phi_j(-q)}\vert_{\pmb{\phi}_{min}} + \int_p \partial_t \phi_{l,min}\:\Gamma^{(3)}_k[q,-q,p,\pmb{\phi}_{min}]_{i,j,l}}
  \end{split}
  \label{eq:dtgamma2}
 \end{equation}
\end{widetext}
The first term of thr r.h.s of Eq.~\eqref{eq:dtgamma2} is obtained from the second functional derivative of Eq.~\eqref{eq:flowGamma} that involves $\Gamma^{(3)}_k$ and $\Gamma^{(4)}_k$ computed for $\pmb{\phi}=\pmb{\phi}_{min}$. With the LPA' {\it ansatz} \eqref{eq:LPA} these quantities are given by the third and fourth derivatives of $U_k$ w.r.t. $\phi_i$.
The calculation is tedious but straighforward and we find :
\begin{equation}
 \partial_t Z_k = 4 \kappa\frac{4\eta_k-d-6}{d+2}  \frac{ (36 \kappa \lambda_6 + u )^2}{(1+m_{k,L}^2)^2(1+m_{k,T}^2)^2},
\end{equation}
from which $\eta_k$ is computed.